\documentclass[twocolumn,showpacs,aps,prb,superscriptaddress,floatfix]{revtex4}

\usepackage{graphicx}
\usepackage{dcolumn}
\usepackage{bm}
\begin{document}
\preprint{}

\title{Field-induced magnetic behavior of the bilayer iridate Sr$_3$Ir$_2$O$_7$}

\author{J.P. Clancy}
\affiliation{Department of Physics, University of Toronto, Toronto, Ontario, M5S 1A7, Canada}

\author{K.W. Plumb}
\affiliation{Department of Physics, University of Toronto, Toronto, Ontario, M5S 1A7, Canada}

\author{C.S. Nelson}
\affiliation{National Synchrotron Light Source, Brookhaven National Laboratory, Upton, New York 11973, USA}

\author{Z. Islam}
\affiliation{Advanced Photon Source, Argonne National Laboratory, Argonne, Illinois 60439, USA}

\author{G. Cao}
\affiliation{Department of Physics and Astronomy, University of Kentucky, Lexington, Kentucky 40506, USA}

\author{T. Qi}
\affiliation{Department of Physics and Astronomy, University of Kentucky, Lexington, Kentucky 40506, USA}

\author{Young-June Kim}
\affiliation{Department of Physics, University of Toronto, Toronto, Ontario, M5S 1A7, Canada}

\begin{abstract}
We have performed resonant magnetic x-ray scattering (RMXS) measurements on single crystal samples of the bilayer iridate Sr$_3$Ir$_2$O$_7$.  We observe the development of antiferromagnetic order below T* $\sim$ 260 K, which persists down to T = 4 K under both field-cooled (fc) and zero-field-cooled (zfc) conditions.  The temperature dependence of the fc and zfc magnetic Bragg peaks suggests the presence of competition between canted and collinear antiferromagnetic ground states.  Under zfc conditions, we observe a suppression of magnetic peak intensity at T$_D$ $\sim$ 50 K which can be attributed to a spin reorientation transition.  Under fc conditions, we find that the canted antiferromagnetic state is stabilized over a much wider range of temperatures, with significantly enhanced magnetic correlation lengths within the Ir-O layers.  The field dependence of the magnetic Bragg peaks provides no evidence of field-induced phase transitions for H $\le$ 4 T.
\end{abstract}


\maketitle

\section{Introduction}
In recent years, the physics of iridium-based 5d transition metal oxides, or iridates, has begun to attract considerable attention\cite{Cao_PRB_2002, Nagai_JPCM_2007, Moon_PRL_2008, Franke_PRB_2011, Matsuhata_JSSC_2004, Subramanian_MRB_1994, Nagai_JLTP_2003, Boseggia_arXiv_2012, JWKim_arXiv_2012, Dhital_arXiv_2012, JHKim_arXiv_2012, Crawford_PRB_1994, Cao_PRB_1998, Kim_PRL_2008, Kim_Science_2009, Jackeli_PRL_2009, Chaloupka_PRL_2010, Singh_PRL_2012, Shitade_PRL_2009, Liu_PRB_2011, Pesin_NP_2010, Yang_PRB_2010, Wan_PRB_2011, Kargarian_PRB_2011, Witczak-Krempa_PRB_2012, Clancy_arXiv_2012, Laguna-Marco_PRL_2010, Cao_PRB_2007, Carter_PRB_2012, Zeb_arXiv_2012}.  Due to the broad, spatially-extended nature of 5d electronic wavefunctions, these materials tend to exhibit a combination of weak electronic correlations, strong crystal field effects, and wide electronic bands.  Furthermore, due to the large atomic mass of iridium, these materials also display strong relativistic spin-orbit coupling effects.  In fact, the energy scale associated with these spin-orbit coupling effects ($\zeta_{5d}$ $\sim$ 0.3-0.5 eV)\cite{Kim_PRL_2008, Laguna-Marco_PRL_2010, Clancy_arXiv_2012} is often comparable to those of the electronic correlations or the crystal electric field.  As a result, the iridates display considerable potential for novel spin-orbit-induced physics, and represent promising candidates for a variety of exotic electronic and magnetic ground states.  In the perovskite iridates Sr$_2$IrO$_4$\cite{Crawford_PRB_1994, Cao_PRB_1998, Moon_PRL_2008, Kim_PRL_2008, Kim_Science_2009, Jackeli_PRL_2009} and Sr$_3$Ir$_2$O$_7$\cite{Cao_PRB_2002, Nagai_JPCM_2007, Moon_PRL_2008, Franke_PRB_2011, Matsuhata_JSSC_2004, Subramanian_MRB_1994, Nagai_JLTP_2003, Boseggia_arXiv_2012, JWKim_arXiv_2012, Dhital_arXiv_2012, JHKim_arXiv_2012}, the combination of modest electronic correlations and strong 5d spin-orbit coupling effects is believed to result in the formation of a j$_{eff}$ = 1/2 spin-orbital Mott insulating state\cite{Kim_Science_2009, Kim_PRL_2008, Boseggia_arXiv_2012, JWKim_arXiv_2012}.  In the honeycomb iridates A$_2$IrO$_3$ (A = Na, Li)\cite{Chaloupka_PRL_2010, Singh_PRL_2012, Shitade_PRL_2009, Liu_PRB_2011}, these effects lead to a potential realization of the Kitaev-Heisenberg model\cite{Chaloupka_PRL_2010, Singh_PRL_2012}, which describes a system of S = 1/2 moments on a honeycomb lattice with highly anisotropic exchange interactions.  In the case of the pyrochlore iridates R$_2$Ir$_2$O$_7$ (R = Y or lanthanide), there are theoretical arguments which predict exotic topological states such as topological insulators, axion insulators, and topological (or Weyl) semi-metals\cite{Pesin_NP_2010, Yang_PRB_2010, Wan_PRB_2011, Kargarian_PRB_2011, Witczak-Krempa_PRB_2012}.

Sr$_3$Ir$_2$O$_7$ is a compound which belongs to the Ruddlesden-Popper series Sr$_{n+1}$Ir$_n$O$_{3n+1}$ [Ref. 5].  The n = 1 member of this series, Sr$_2$IrO$_4$, is a single-layered perovskite which is isostructural to La$_2$CuO$_4$, the parent compound for the high-T$_C$ cuprate superconductors\cite{Crawford_PRB_1994}.  Sr$_2$IrO$_4$ is an antiferromagnetic insulator, which displays strongly anisotropic quasi-two-dimensional physical properties\cite{Cao_PRB_1998, Moon_PRL_2008}.  In the opposite limit (n = $\infty$), the orthorhombic perovskite SrIrO$_3$ is a correlated metal which displays effectively three-dimensional physical properties\cite{Cao_PRB_2007, Moon_PRL_2008, Zeb_arXiv_2012}.  As the n = 2 member of this series, the bilayered perovskite Sr$_3$Ir$_2$O$_7$ displays properties which lie in the intermediate region between these two compounds.  While Sr$_3$Ir$_2$O$_7$ is an insulator, its low resistivity suggests that it falls close to the metal-insulator boundary\cite{Cao_PRB_2002, Nagai_JPCM_2007, Moon_PRL_2008}.  Similarly, while the system exhibits anisotropic magnetization and transport properties\cite{Cao_PRB_2002}, its bilayer character implies that interlayer couplings are significant and may play a important role in shaping the physics of the system.

Sr$_3$Ir$_2$O$_7$ has a layered perovskite structure which consists of strongly coupled bilayers of IrO$_6$ octahedra\cite{Cao_PRB_2002, Subramanian_MRB_1994, Matsuhata_JSSC_2004}.  These bilayers are stacked along the crystallographic {\it c}-axis, with pairs of adjacent bilayers separated by an interlayer of Sr-O and offset by half a unit cell in the {\it ab}-plane.  The spacing between the Ir-O layers in adjacent bilayers is approximately 6.43 {\AA}, while the spacing between Ir-O layers within the same bilayer is 4.07 {\AA} [Refs. 1,5].  This can be compared with the structure of the single-layered perovskite Sr$_2$IrO$_4$, in which the spacing between Ir-O layers is a uniform 6.48 {\AA} [Ref. 12].  The IrO$_6$ octahedra within each bilayer are elongated along the {\it c}-axis, giving rise to two long Ir-O bonds (2.035 {\AA}) and four short Ir-O bonds (1.989 {\AA}) per octahedra\cite{Cao_PRB_2002}.  Neighboring octahedra are also rotated about the {\it c}-axis by an angle of $\sim$ 12$^{\circ}$, resulting in a staggered structure in which the Ir-O planes within each bilayer are slightly out of phase with each other\cite{Cao_PRB_2002, Subramanian_MRB_1994}.  The rotation of the IrO$_6$ octahedra is particularly important because it breaks the inversion symmetry between nearest-neighbor Ir sites, leading to the development of significant Dzyaloshinskii-Moriya interactions.

The magnetism in this material arises from the magnetic moments carried by Ir$^{4+}$ ions (5d$^5$ electronic configuration).  The octahedral crystal field environment at the Ir site splits the 5d electronic states into a lower t$_{2g}$ manifold and an upper e$_g$ manifold, and the splitting between these levels is sufficiently large that the Ir ions adopt a low-spin S = 1/2 ground state.  However, due to the presence of strong 5d spin-orbit coupling effects, the low-lying t$_{2g}$ manifold is further split into a four-fold degenerate j$_{eff}$ = 3/2 band and a two-fold degenerate j$_{eff}$ = 1/2 band.  The lower-lying j$_{eff}$ = 3/2 levels become fully occupied, while the j$_{eff}$ = 1/2 band remains half-filled.  From this point, even the relatively weak 5d electronic correlations are sufficient to split the j$_{eff}$ = 1/2 states into a lower Hubbard band and an  upper Hubbard band, and this leads to the formation of a small insulating gap\cite{Moon_PRL_2008,Kim_PRL_2008, Kim_Science_2009}.

Bulk magnetization measurements on Sr$_3$Ir$_2$O$_7$ reveal an anomaly at T$_{C}$ $\sim$ 285 K which has been attributed to the development of a weak ferromagnetic or canted antiferromagnetic ground state\cite{Cao_PRB_2002, Nagai_JPCM_2007, Nagai_JLTP_2003}.  The ordered moment associated with this magnetic ground state appears to be very small\cite{Cao_PRB_2002, Nagai_JPCM_2007, Nagai_JLTP_2003}, and a saturation moment of $\mu_s$ = 0.037 $\mu_B$/Ir has been observed within the {\it ab}-plane\cite{Cao_PRB_2002}.  This is over 25 times smaller than the value of $\mu_s$ = 1 $\mu_B$/Ir which is expected for a localized S = 1/2 moment.  The magnetization of Sr$_3$Ir$_2$O$_7$ is also clearly anisotropic, with the in-plane magnetization (M$_{ab}$) greater than the magnetization along the stacking direction (M$_c$) by a factor of 2 to 4 [Ref. 1].  Perhaps most intriguingly, the magnetization of this compound appears to be very sensitive to the presence of applied magnetic fields, displaying a surprisingly rich H-T phase upon field-cooling.  In addition to the transition observed at T$_C$, field-cooling in a modest field (0 $\le$ H$_{ab}$ $\le$ 0.25 T) introduces two new magnetic anomalies at T* $\sim$ 260 K and T$_D$ $\sim$ 50 K [Ref. 1].  At T*, there is a noticeable change in the slope of M$_{ab}$(T), and the magnetization begins to increase more quickly with decreasing temperature.  Below T$_D$, M$_{ab}$(T) begins to decrease quite rapidly, undergoing a reversal in sign below T $\sim$ 20 K.  Field-cooling in the presence of higher fields (H $\ge$ 0.25 T) suppresses the anomaly at T$_D$, replacing the downturn in M$_{ab}$ with a sharp upturn, and eliminating any evidence of a magnetization reversal.  Interestingly, the temperature scale associated with the field-cooled magnetization reversal also appears to correspond to a magnetic anomaly observed in zero-field-cooled $\mu$SR experiments\cite{Franke_PRB_2011}.  The muon precession signal in Sr$_3$Ir$_2$O$_7$ undergoes a qualitative change below T $\sim$ 25 K, suggesting some form of change in the ordered spin structure at low temperatures. 

The magnetic structure of Sr$_3$Ir$_2$O$_7$ has recently been investigated using both resonant magnetic x-ray scattering\cite{Boseggia_arXiv_2012, JWKim_arXiv_2012} and conventional neutron scattering techniques\cite{Dhital_arXiv_2012}.  These studies suggest that the magnetically ordered state which develops below T$_C$ has a G-type antiferromagnetic (AF) structure, which can be characterized by an ordering wave-vector of {\bf q} = (1/2, 1/2, 0).  This structure consists of magnetic moments which are antiferromagnetically coupled to all nearest neighbors, both within the Ir-O layers and along the {\it c}-axis stacking direction.  The detailed temperature and field dependence of this magnetically ordered structure remain largely unexplored, although the development of magnetic Bragg peaks has been reported at transition temperatures ranging from 230 K [Ref. 8] to 280 K [Ref. 9].  Interestingly, x-ray polarization analysis of the magnetic signal suggests that the direction of the ordered moments in Sr$_3$Ir$_2$O$_7$ is parallel to the crystallographic {\it c}-axis\cite{JWKim_arXiv_2012}.  This is somewhat surprising, as the magnetic easy-axis in the single-layer compound Sr$_2$IrO$_4$ is reported to lie within the {\it ab}-plane\cite{Kim_Science_2009}.  It has been argued that this 90$^{\circ}$ rotation of the spin orientation may be driven by the much stronger interlayer coupling effects present within the bilayer crystal structure\cite{JWKim_arXiv_2012}.  The resonant enhancement of the magnetic Bragg peaks at the Ir L$_2$ and L$_3$ absorption edges is found to differ quite dramatically in this compound, with observed peak intensities at the L$_3$ edge being larger by a factor of 30 [Ref. 8] to 100 [Ref. 9].  As in the case of Sr$_2$IrO$_4$\cite{Kim_Science_2009}, this remarkably large L$_3$/L$_2$ intensity ratio is taken as evidence to support the existence of a j$_{eff}$ = 1/2 spin-orbital Mott insulating state\cite{Boseggia_arXiv_2012, JWKim_arXiv_2012}.

In this article, we report an investigation of the field-induced magnetic properties of Sr$_3$Ir$_2$O$_7$ using resonant magnetic x-ray scattering (RMXS) techniques.  In particular, we have studied how the magnetic Bragg peaks in this compound evolve as a function of temperature and field, performing detailed RMXS measurements under both zero-field-cooled (zfc) and field-cooled (fc) conditions.  Our results suggest that the H-T phase diagram of Sr$_3$Ir$_2$O$_7$ can be understood in terms of competition between canted antiferromagnetic and collinear antiferromagnetic ground states, with potential spin reorientation transitions driven by temperature and field-cooling.  Under zfc conditions the system appears to favor a collinear antiferromagnetic state, while under fc conditions the canted antiferromagnetic state is preferred.  These results clearly demonstrate that field-cooling, even in the presence of extremely small magnetic fields (H $\le$ 0.015 T), can have a profound influence on the magnetic properties of this system.  Interestingly, in spite of this sensitivity to fc effects, the application of higher magnetic fields appears to have almost no impact on the behavior of Sr$_3$Ir$_2$O$_7$.  In contrast to the single-layered compound Sr$_2$IrO$_4$, we find no evidence of field-induced magnetic phase transitions for applied fields of up to H = 4 T.

\section{Experimental Details}

Single crystal samples of Sr$_3$Ir$_2$O$_7$ were prepared using self-flux techniques, as described elsewhere\cite{Cao_PRB_2002}.  Sr$_3$Ir$_2$O$_7$ exhibits a tendency to grow as thin plate-like single crystals, with the crystallographic {\it c}-axis normal to the sample surface.  Detailed magnetization measurements on these samples have previously been reported by Cao et al\cite{Cao_PRB_2002}.  For illustrative purposes, representative zero-field-cooled (zfc) and field-cooled (fc) magnetization curves have been provided in Figure 3(a).  The sample used in this experiment had dimensions of approximately 1.0 $\times$ 1.0 $\times$ 0.1 mm$^3$.  The mosaicity of the sample was determined by a series of x-ray rocking scans.  Although the full mosaic spread was found to be relatively large ($\sim$ 2 degrees), the dominant grain had a full-width at half-maximum (FWHM) of only 0.11 degrees and a scattering intensity which was more than an order of magnitude stronger than any of the secondary grains.

Resonant magnetic x-ray scattering measurements were performed using Beamline X21 at the National Synchrotron Light Source (NSLS) at Brookhaven National Laboratory, and Beamline 6-ID-B at the Advanced Photon Source (APS) at Argonne National Laboratory.  In both cases, measurements were carried out at the Ir L$_{3}$ absorption edge (E = 11.215 keV).  This absorption edge probes transitions from the occupied Ir 2p$_{3/2}$ states to the unoccupied 5d$_{3/2}$ and 5d$_{5/2}$ states, and provides a large resonant enhancement to the cross section for magnetic x-ray scattering.  

For measurements on X21, the sample was mounted on the coldfinger of a closed cycle cryostat and aligned within a Huber four-circle diffractometer.  The base temperature for this sample environment was T = 7 K.  Measurements were performed using vertical scattering geometry, wherein the polarization of the incident beam is perpendicular to the scattering plane defined by {\bf k$_i$} and {\bf k$_f$} (i.e. $\sigma$ polarization).  The incident x-ray beam was monochromated using a Si-(111) double crystal monochromator.  Although a LiF-(200) analyzer was employed to improve resolution and reduce background, no polarization analysis was performed.  As a result, the X21 measurements are sensitive to the total scattering intensity, which includes contributions from both the rotated ($\sigma$-$\pi$) and unrotated ($\sigma$-$\sigma$) polarization channels.  Resonant x-ray scattering measurements were performed on the sample under both field-cooled and zero-field-cooled conditions.  Field-cooling was achieved using a small permanent magnet assembly capable of generating a magnetic field of H = 0.015T at the sample position.  The direction of the applied field was in the {\it ab} plane, at an angle of $\sim$ 22.5$^{\circ}$ with respect to the [1$\overline{1}$0] axis.  To maximize the effect of field-cooling on the magnetization of the sample\cite{Cao_PRB_2002}, the field was applied at T $\sim$ 300 K, well above T$_C$.

For measurements on 6-ID-B, the sample was mounted in a vertical field cryomagnet with a base temperature of T = 4K and a maximum field of H = 4 T.  Measurements were performed using horizontal scattering geometry, wherein the polarization of the incident beam is parallel to the scattering plane defined by {\bf k$_i$} and {\bf k$_f$} (i.e. $\pi$ polarization).  The incident x-ray beam was monochromated by a Si-(111) double crystal monochromator.  Polarization analysis was performed using the (3, 3, 3) reflection from an Al-(111) analyzer.  In this configuration, the scattering term with a rotated polarization vector (i.e. the $\pi$-$\sigma$ channel) is magnetic in origin, and has a scattering intensity proportional to {\bf k$_f$} $\cdot$ {\bf M} [Ref. 31].  Alternatively, the term which possesses an unrotated polarization vector (i.e. the $\pi$-$\pi$ channel) includes contributions from both charge scattering (proportional to {\bf k$_f$} $\cdot$ {\bf k$_i$}) and magnetic scattering (proportional to [{\bf k$_f$} $\times$ {\bf k$_i$}] $\cdot$ {\bf M}) [Ref. 31].  The sample was aligned such that the (H, H, L) plane was coincident with the horizontal scattering plane.  In this orientation, the direction of the applied field was parallel to the [1$\overline{1}$0] axis.

\section{Experimental Results}
\subsection{Characterization of Magnetic Bragg Peaks}

\begin{figure}
\includegraphics{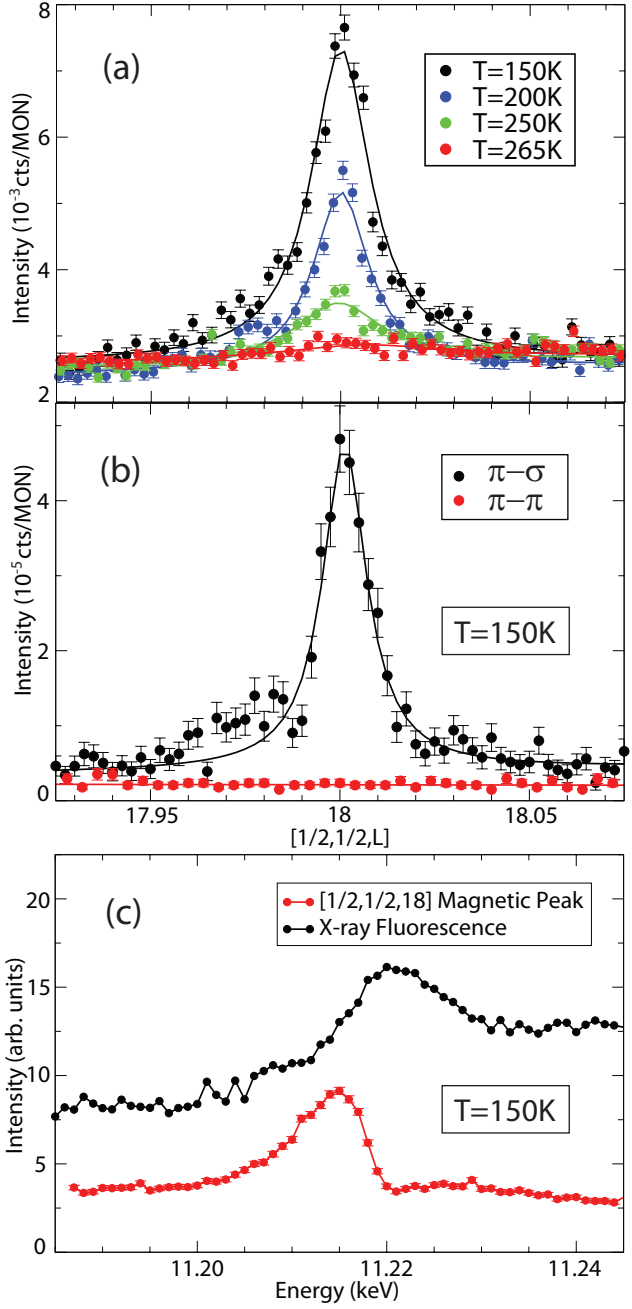}
\caption{(Color online) (a) Temperature dependence of the (1/2, 1/2, 18) magnetic Bragg peak under zero-field-cooled (zfc) conditions.  The total scattering intensity includes both rotated ($\sigma$-$\pi$) and unrotated ($\sigma$-$\sigma$) scattering contributions.  (b) Polarization dependence of the (1/2, 1/2, 18) magnetic Bragg peak under zfc conditions.  The magnetic scattering contribution can be observed in the rotated ($\pi$-$\sigma$) polarization channel.  (c) Energy dependence of the (1/2, 1/2, 18) magnetic Bragg peak at the Ir L$_3$ absorption edge under zfc conditions.  X-ray fluorescence measurements are provided for the purpose of energy calibration.}
\end{figure}

A series of representative RMXS measurements are provided in Figure 1.  These measurements have been carried out at the (1/2, 1/2, 18) position in reciprocal space, which is an antiferromagnetic ordering wave-vector consistent with the magnetic structure previously proposed for this material\cite{Boseggia_arXiv_2012, Dhital_arXiv_2012, JWKim_arXiv_2012}.  For convenience, we have adopted the same tetragonal unit cell notation ({\it a} = {\it b} = 3.90 {\AA}, {\it c} = 20.86 {\AA}) as employed in Reference 8.  However, it should be noted that the true space group of Sr$_3$Ir$_2$O$_7$ is still the subject of some debate.  Several different space groups, including the tetragonal I4/mmm\cite{Subramanian_MRB_1994} and the orthorhombic Bbca [Ref. 1] and Bbcb [Ref. 6], have been reported for this compound in the literature.  These space groups are chiefly distinguished by the nature of the rotations associated with the IrO$_6$ octahdedra.  In the event that these octahedral rotations are correlated, the resulting space group symmetry will be orthorhombic, whereas in the case that they are not the space group will be tetragonal.  It has been suggested that the nature of these octahedral rotations may be sensitive to sample disorder or deviations from ideal oxygen stoichiometry\cite{Boseggia_arXiv_2012}, however this has yet to be investigated in systematic detail.  X-ray structure refinements performed on Sr$_3$Ir$_2$O$_7$ crystals from the same growth batch as our sample are consistent with the Bbca (correlated rotation) space group\cite{Cao_PRB_2002}.

The data provided in Figure 1 serves to verify that the (1/2, 1/2, 18) Bragg peak exhibits the (a) temperature dependence, (b) polarization dependence, and (c) energy dependence, that should be associated with a magnetic feature.  All of the measurements presented in this figure have been collected under zfc conditions, with the data in panels (a) and (c) obtained at X21 and the data in panel (b) obtained at 6-ID-B.  Figure 1(a) consists of a series of reciprocal space scans taken along the [1/2, 1/2, L] direction at temperatures ranging from 150 K (T$_D$ $\ll$ T $\ll$ T*) to 265 K (T* $<$ T $<$ T$_C$).  The intensity of the (1/2, 1/2, 18) peak decreases monotonically upon warming, eventually disappearing at T $\sim$ T*.  While a small peak is observable below T* at T = 250 K, there is no peak visible in the intermediate state between T* = 260 K and T$_C$ = 285 K.  A more detailed study of the magnetic temperature dependence is provided in Figure 3, and will be discussed further in Section IIIB.  Figure 1(b) consists of [1/2, 1/2, L] scans which have been carried out using polarization analysis to distinguish between the rotated ($\pi$-$\sigma$ polarization channel) and unrotated ($\pi$-$\pi$ polarization channel) scattering contributions.  These scans show that the scattering observed at (1/2, 1/2, 18) occurs exclusively in the $\pi$-$\sigma$ channel, which is consistent with the polarization dependence expected for magnetic scattering with an ordered moment aligned along the {\it c}-axis.  Figure 1(c) illustrates the energy dependence of the scattering at (1/2, 1/2, 18), plotting the intensity of the peak as a function of the incident energy of the x-ray beam.  As the incident energy is scanned through the Ir L$_3$ edge, the feature at (1/2, 1/2, 18) exhibits clear resonant behavior, displaying a well-defined peak in energy at E$_i$ $\sim$ 11.215 keV.  This can be compared with the energy dependence of the x-ray fluorescence signal, as measured by an energy-sensitive detector perpendicular to the incident beam.

\begin{figure}
\includegraphics{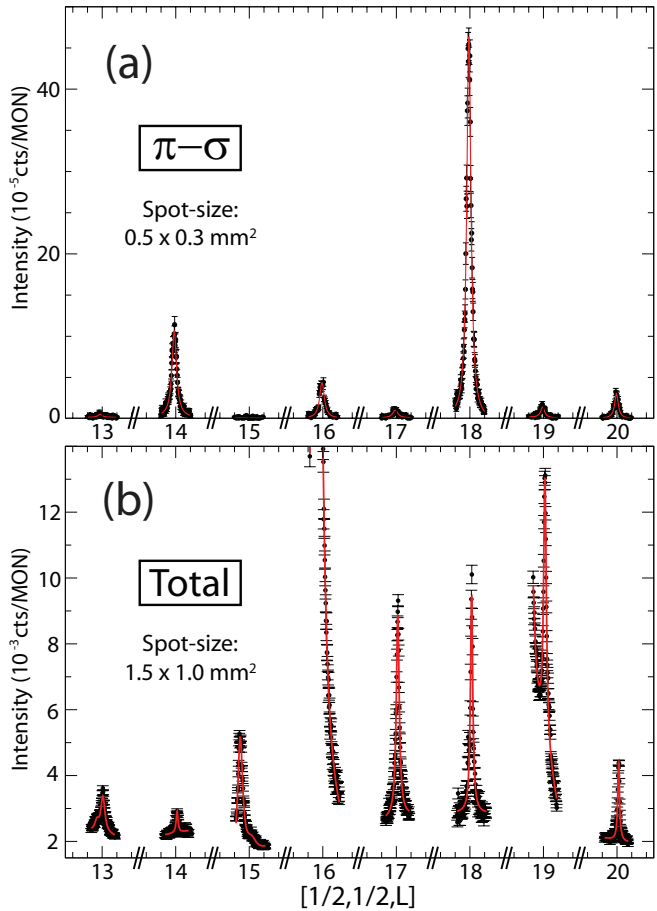}
\caption{(Color online) A survey of magnetic Bragg peaks along the [1/2, 1/2, L] direction.  (a) Measurements performed on 6-ID-B using a focused spot-size and $\pi$-$\sigma$ polarization analysis to isolate the magnetic contribution to the scattering intensity.  These measurements were collected under field-cooled conditions (H = 0.1 T) at T = 150 K.  (b) Measurements performed on X21 using a broad spot-size and no polarization analysis.  These measurements were collected under zero-field-cooled conditions (H = 0 T) at T = 7 K.  Note that strong background powder lines are produced near L = 16 and L = 19 due to scattering from the sample mount.  Each scan extends $\pm$0.1 r.l.u. around the Bragg position.}
\end{figure}

A survey of magnetic Bragg peaks along the [1/2, 1/2, L] direction is provided in Figure 2.  The data presented in Figure 2(a) was obtained at 6-ID-B, while the data in Figure 2(b) was obtained at X21.  The 6-ID-B data set was collected at T = 150 K, under fc conditions with H = 0.1 T.  These measurements utilized a polarization analyzer to isolate the magnetic scattering contribution, and employed a focused x-ray beam with a spot size of approximately 0.5 mm $\times$ 0.3 mm (horizontal $\times$ vertical).  With the exception of L = 15, this data set provides evidence of magnetic Bragg peaks at all (1/2, 1/2, L) positions measured.  For L = even, these magnetic peaks are quite strong, while for L = odd they appear considerably weaker.  This observation is consistent with the proposed magnetic structure for Sr$_3$Ir$_2$O$_7$ which was recently put forward by Boseggia et al\cite{Boseggia_arXiv_2012}.  As discussed in Section I, this magnetic structure consists of antiferromagnetically ordered spins with an ordering wave-vector of {\bf q} = (1/2, 1/2, 0).  However, due to the bilayer structure of Sr$_3$Ir$_2$O$_7$, this ordering wave-vector can give rise to two different propagation vectors depending on the order of the magnetic stacking sequence: {\bf k$_a$} = (1/2, 1/2, 0) and {\bf k$_b$} = (1/2, -1/2, 0).  Domains which are described by the {\bf k$_a$} propagation vector will result in magnetic peaks at (1/2, 1/2, odd) positions, while domains described by {\bf k$_b$} will produce peaks at the (1/2, 1/2, even) positions.  Based on the intensity of the L = odd and L = even reflections observed in Figure 2(a), this data set is consistent with domains which are predominantly described by the {\bf k$_b$} propagation vector.  Real space RMXS domain mapping suggests that the dimensions of these magnetic domains are on the order of $\sim$ 100 $\mu$m by 100 $\mu$m [Ref. 8], so the beam size on 6-ID-B is sufficiently large to integrate over a small number of distinct grains.  The existence of this domain structure can be verified, at least in part, by comparing data sets collected with varying spot sizes.  Such a comparison is provided by the data in Figure 2(b), which shows measurements performed on X21 with a significantly larger spot size (approximately 1.5 mm $\times$ 1.0 mm) and no polarization analysis.  The strong background features which arise near L = 16 and L = 19 are associated with powder lines caused by scattering off the copper sample mount.  The remaining (1/2, 1/2, L) peaks are all magnetic in nature, and clearly display a much more balanced distribution of scattering intensity for even/odd L than in Figure 2(a).  This result makes intuitive sense if one considers that the larger beam size will effectively integrate over a much larger number of magnetic domains on the sample surface.

Within the limits of the experimental uncertainty, we find that the magnetic structure along [1/2, 1/2, L] appears to be almost identical under zero-field-cooled and field-cooled conditions (with H = 0.015 T or H = 0.1 T).  Similarly, while there are global variations in the intensity of the magnetic Bragg peaks between T = 150 K and T = 5 K, the form of the magnetic structure factor itself does not appear to exhibit any significant temperature dependence.  This observation seems to be consistent with the results of other studies on Sr$_3$Ir$_2$O$_7$ performed using both x-rays\cite{Boseggia_arXiv_2012, JWKim_arXiv_2012} and neutrons\cite{Dhital_arXiv_2012}.

\subsection{Temperature Dependence of Magnetic Order}

The temperature dependence of the (1/2, 1/2, 18) magnetic Bragg peak is presented in Figure 3(b).  This figure plots the integrated intensity of the (1/2, 1/2, 18) peak as a function of temperature under both zfc (H = 0 T) and fc (H = 0.015 T) conditions.  No polarization analysis has been employed for these measurements, so the integrated intensity includes contributions from both the $\sigma$-$\sigma$ and $\sigma$-$\pi$ polarization channels.  To enable a direct comparison of magnetic properties, the temperature dependence of the in-plane magnetization for this compound, M$_{ab}$, has been provided in Figure 3(a).  As noted in Section I, the zfc magnetization of Sr$_3$Ir$_2$O$_7$ displays a rather weak magnetic anomaly at T$_{C}$ = 285 K, while the fc magnetization exhibits a series of anomalies at T$_C$ = 285 K, T* $\sim$ 260 K, and T$_D$ $\sim$ 50 K.  The most dramatic feature in M$_{ab}$(T), the field-induced magnetization reversal, occurs at T $\sim$ 20 K.  

\begin{figure}
\includegraphics{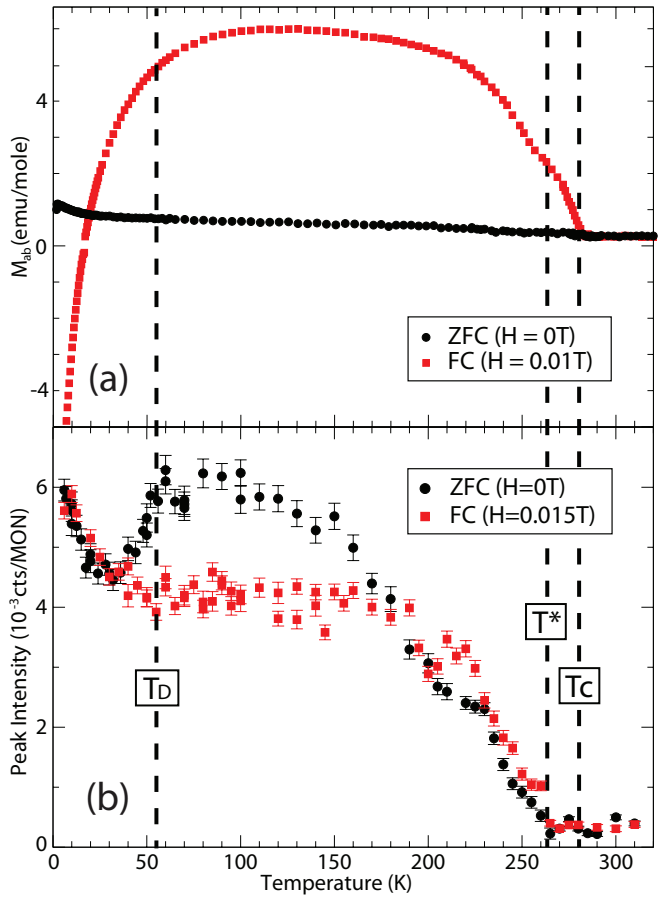}
\caption{(Color online) (a) Magnetization of Sr$_3$Ir$_2$O$_7$ under zero-field-cooled (H = 0 T) and field-cooled (H = 0.01T) conditions.  All magnetization measurements were carried out with a field of H = 0.01 T applied within the {\it ab}-plane.  (b) Temperature dependence of the (1/2, 1/2, 18) magnetic Bragg peak under zero-field-cooled (H = 0 T) and field-cooled (H = 0.015T) conditions.  No polarization analysis has been employed for these measurements.  The direction of the applied field is within the {\it ab}-plane, at an angle of 22.5$^{\circ}$ with respect to [1$\bar{1}$0].}
\end{figure}

The zfc x-ray data (represented by the black circles in Figure 3(b)) indicates that the first magnetic scattering intensity at (1/2, 1/2, 18) develops at approximately T $\sim$ 260 K.  Interestingly, while this transition temperature agrees very well with the fc anomaly at T*, it falls significantly below the zfc transition at T$_C$ = 285 K.  This may indicate that the intermediate state between T* and T$_C$ is characterized by a different magnetic structure, which is perhaps incommensurate or short-range-ordered in nature.  While this discrepancy is somewhat surprising, it should be noted that previous x-ray studies on Sr$_3$Ir$_2$O$_7$\cite{Boseggia_arXiv_2012} have reported even larger differences in temperature scale (T$_C$ $\sim$ 275 K, T$_{Bragg}$ = 230 K).  A second unusual feature can be observed in the zfc data at T $\sim$ 50 K, whereupon the magnetic peak intensity abruptly drops (from $\sim$ 50 K to 30 K) and then recovers to its initial value (from $\sim$ 30 K to 5 K).  This anomalous suppression of magnetic peak intensity is highly reproduceable, and can be observed on both warming and cooling with no apparent hysteresis.  It is worth noting that the onset of this zfc suppression begins in the vicinity of the fc anomaly T$_D$, and the intensity minimum closely corresponds to the zfc $\mu$SR anomaly reported at T $\sim$ 25 K [Ref. 7].

The fc x-ray data (represented by the red squares in Figure 3(b)) indicates that field-cooling, even in the presence of extremely small magnetic fields, has a significant impact on the evolution of the magnetic peak intensity.  As in the zfc data, the fc magnetic Bragg peaks first appear at T $\sim$ T*, and steadily grow in intensity with decreasing temperature.  However, the fc and zfc curves begin to diverge just below 200 K, and the fc peak intensity appears to saturate at approximately 2/3 of the zfc value.  Below T$_D$, the intensity of the fc magnetic peak begins to increase again, leading to a reconnection of the fc and zfc curves at T $\sim$ 30 K.

\subsection{Field Dependence of Magnetic Order}

The field dependence of the (1/2, 1/2, 18) magnetic Bragg peak is illustrated by the representative data sets provided in Figure 4.  These data sets were collected at T = 5K, under zfc (H = 0 T) and fc (H = 0.1 T, 0.5 T) conditions.  These measurements were carried out on 6-ID-B, employing polarization analysis to isolate the magnetic scattering contribution.  To enable an accurate comparison of the zfc and fc peak widths, all data sets have been normalized to unit scattering intensity.  The field dependence of the peak widths along the interlayer stacking direction, [0,0,L], is provided in Figure 4(a).  Note that the [0,0,L] peak width is significantly broader than the resolution limit, indicating the presence of finite magnetic correlation lengths along the {\it c}-axis stacking direction.  These finite correlation lengths appear to be unaffected by field-cooling or by applied fields of up to H = 0.5 T.  In contrast, the correlation lengths along the [H,H,0] direction appear to be significantly enhanced by field-cooling.  The field dependence of the [H,H,0] peak width, which reflects the magnetic correlations within the Ir-O planes, is provided in Figure 4(b).  Note that the [H,H,0] peak width decreases substantially upon field-cooling, and becomes almost resolution-limited at H = 0.1 T.  The application of higher magnetic fields (H = 0.5 T) results in a slight additional narrowing of the magnetic Bragg peak, giving rise to a fully resolution-limited lineshape and correlation lengths consistent with long-range magnetic order.

The field dependence of the correlation lengths can be investigated in a more quantitative fashion by performing simple Lorentzian fits to the data sets provided in Figure 5.  These fits can be used to extract a value for the FWHM of the magnetic peak, $\Gamma_{obs}$.  This observed peak width is related to the intrinsic width of the magnetic peak by the expression $\Gamma_{int}$ = $\sqrt{\Gamma_{obs}^2-\Gamma_{res}^2}$, where $\Gamma_{res}$ refers to the FWHM of the experimental resolution function.  In this case, $\Gamma_{res}$ can be determined from the FWHM of a nearby structural Bragg peak.  The relevant magnetic correlation lengths can then be extracted from the expression $\xi$ = [(2$\pi$/d)($\Gamma_{int}$/2)]$^{-1}$.  Along the interlayer stacking direction, we find correlation lengths of $\sim$ 200 {\AA}.  This length scale is equivalent to approximately 10 unit cells, or 20 Ir-O bilayers.  Within the Ir-O planes, the correlation lengths are much longer, ranging from $\sim$ 1000 {\AA} (for H = 0 T) to 2700 {\AA} (for H = 0.1 T) to effectively infinite (for the resolution-limited H = 0.5 T data set).  This difference in correlation lengths along the [H,H,0] and [0,0,L] directions may reflect the layered quasi-2D structure of Sr$_3$Ir$_2$O$_7$, which naturally favors longer and better defined correlation lengths within the strongly coupled Ir-O layers.  Alternatively, it may reflect the presence of defects in the stacking sequence, which are known to occur quite commonly in many stacked bilayer compounds.

Given the significant impact of field-cooling on the phase diagram of Sr$_3$Ir$_2$O$_7$ at low fields, it is only natural to investigate the effects of higher fields on the system as well.  In the single-layered iridate Sr$_2$IrO$_4$, a field-induced magnetic ``spin-flop'' transition has been reported at H$_C$ = 0.2 T [Ref. 15].  This transition is characterized by a change in the stacking sequence of the canted AF magnetic structure, with the canted moments rotating to align the net magnetization of each layer with the direction of the applied field.  Due to the apparent similarities between the antiferromagnetic ground states of Sr$_3$Ir$_2$O$_7$ and Sr$_2$IrO$_4$ there is strong motivation to believe that similar spin-flop transitions could also occur in this compound.  In addition, magnetization measurements on Sr$_3$Ir$_2$O$_7$ suggest that there is some significance to the energy scale of H $\sim$ 0.25 T, as this is magnitude of the the field required to suppress the magnetization drop below T$_D$ [Ref. 1].

\begin{figure}
\includegraphics{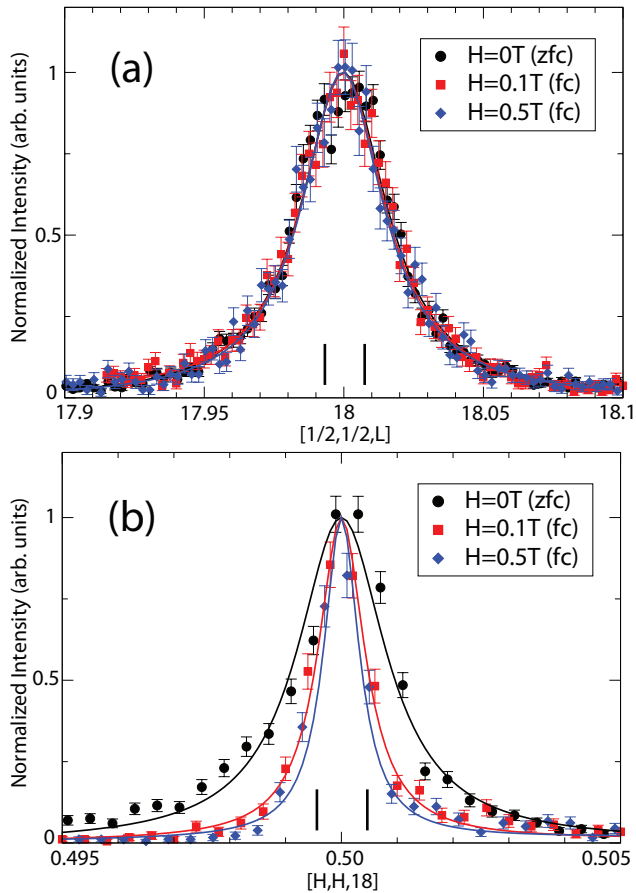}
\caption{(Color online) The effect of applied field on the (1/2, 1/2, 18) magnetic Bragg peak.  Representative scans performed along (a) the [0, 0, L] direction, and (b) the [H, H, 0] direction, are provided for H = 0 T, 0.1 T, and 0.5 T.  The H = 0 T scans were performed after cooling in zero-field, while the H = 0.1 T and 0.5 T scans were performed after field-cooling with H = 0.1 T.  All scans were collected at T = 5 K using $\pi$-$\sigma$ polarization analysis.  The direction of the applied field is parallel to [1$\bar{1}$0].  Note that field-cooling results in a significant decrease in peak width along the [H, H, 0] direction, while the width along [0, 0, L] remains completely unaffected.  The experimental resolution limit is denoted by the vertical black bars.  For purposes of illustration, all scans have been normalized to unit intensity.}
\end{figure}

The field dependence of the (1/2, 1/2, 18) magnetic Bragg peaks at higher fields is provided in Figure 5.  As in the low field measurements, the applied magnetic field is oriented along the [1$\bar{1}$0] direction, perpendicular to the {\it c}-axis stacking direction.  Figure 5(a) shows the intensity of the (1/2, 1/2, 18) magnetic Bragg peak as a function of increasing magnetic field under both zfc and fc conditions.  Note that in the event of a field-induced magnetic phase transition, one would expect to observe a change in the intensity of the magnetic Bragg peaks.  This holds true for a spin-flop transition, as in Sr$_2$IrO$_4$, but also for a more general change in magnetic structure, or even a gradual polarization of the ordered moments.  In horizontal scattering geometry, for the $\pi$-$\sigma$ polarization channel, the resonant magnetic x-ray scattering cross-section is proportional to {\bf M} $\cdot$ {\bf k$_i$} [Ref. 31].  As the direction of the applied magnetic field is perpendicular to {\bf k$_i$}, if the moments align with the direction of the field then the intensity of the magnetic Bragg peak should decrease.  In contrast, the intensity of the (1/2, 1/2, 18) peak remains constant from H = 0 T to H = 4 T under both fc and zfc conditions.  Figure 5(b) provides representative [1/2,1/2,L] scans collected at the H = 0 T and H = 4 T end-points for the field-scans presented in Figure 5(a).  The lack of change in the magnetic peak intensity suggests that there is no difference in the magnetic structure at H = 4 T, while the absence of any change in peak width indicates that there is no difference in magnetic correlation lengths along the interlayer stacking direction.  In addition, exploratory survey measurements at H = 4 T revealed no evidence of new magnetic Bragg peaks forming at previously forbidden positions.  We conclude that there are no high-field magnetic phase transitions in Sr$_3$Ir$_2$O$_7$ for applied fields of up to H = 4 T. 

\begin{figure}
\includegraphics{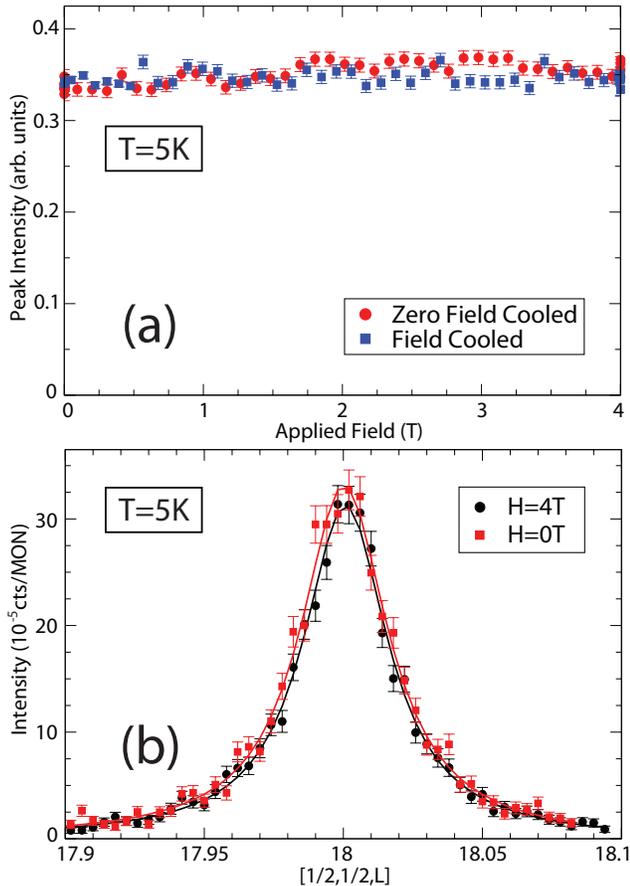}
\caption{(Color online) Field dependence of the (1/2, 1/2, 18) magnetic Bragg peak. (a) Peak intensity as a function of applied magnetic field under field-cooled (H = 0.1 T) and zero-field-cooled (H = 0 T) conditions.  (b) Representative scans performed along the [0, 0, L] direction at H = 0 T and H = 4 T.  All measurements shown here were collected at T = 5 K using $\pi$-$\sigma$ polarization analysis.  The direction of the applied field is parallel to [1$\bar{1}$0].}
\end{figure}

\section{Discussion}

In order to understand the temperature and field dependence of the magnetic Bragg peaks, it is useful to consider the behavior of the antiferromagnetic order parameter, m$_{AF}$.  This order parameter, which is equivalent to the staggered magnetization of the system, is proportional to the square root of the scattering intensity observed at the antiferromagnetic ordering wave-vector (i.e. I $\propto$ m$_{AF}$$^2$).  As a result, the magnetic peak intensities displayed in Figure 3(b) can be taken as a direct measure of the antiferromagnetically-aligned component of the ordered moments.  This can be compared to the magnetization measurements presented in Figure 3(a), which provide a measure of the ferromagnetically-aligned component of the ordered moment.  Recent theoretical calculations\cite{JWKim_arXiv_2012, Private_Communication} suggest that there are two potential magnetic ground states for Sr$_3$Ir$_2$O$_7$ which possess almost equivalent ground state energies.  These ground states correspond to G-type AF orderings with either a collinear AF spin configuration (magnetic easy-axis parallel to the {\it c}-axis stacking direction) or a canted AF configuration (magnetic easy-axis within the {\it ab}-plane).  Each of these spin configurations gives rise to a different value of m$_{AF}$, with the larger m$_{AF}$ corresponding to the collinear AF state (since m$_{AF}$ will be maximized when the ordered spins are antiparallel), and the smaller m$_{AF}$ corresponding to the canted AF state (since m$_{AF}$ will be reduced by a factor of cos$\theta$ when the spins are canted by an angle $\theta$ with respect to the magnetic easy-axis).  While the value of $\theta$ for the canted AF state has not yet been determined experimentally, theoretical calculations suggest that it should be approximately equal to the angle of the octahedral rotations ($\theta$ $\sim$ 12$^{\circ}$)\cite{Private_Communication}.

Applying these ideas to the data in Figure 3(b), it follows that the zfc magnetic ground state can be associated with the collinear AF spin configuration, while the fc ground state can be associated with the canted AF configuration.  As these states are very close in energy, even small perturbations may be sufficient to drive the system from one state into another.  This would explain the anomalous suppression of the zfc peak intensity below T$_D$ $\sim$ 50 K, as well as the sudden increase in fc and zfc peak intensity below T $\sim$ 30 K.  These transitions can both be interpreted as spin reorientation transitions, resulting from changes in the canting angle of the ordered moments.  In the first case, the zfc anomaly at T$_D$ $\sim$ 50 K represents a transition from collinear AF to canted AF.  In the second, the fc/zfc anomaly at T $\sim$ 30 K represents a transition from canted AF to collinear AF.  

If we assume that the magnetic easy-axis in the canted AF state is given by the $\langle$110$\rangle$ direction closest to the applied field, and the canting angle is given by $\theta$ $\sim$ 12$^{\circ}$, then it is possible to calculate a theoretical intensity ratio for the two magnetic ground states.  Since the direction of the magnetic easy-axis rotates by 90$^{\circ}$ between the collinear and canted AF states, this intensity ratio must incorporate two multiplicative factors: (1) a factor of cos$\theta$ due to the change in canting angle, and (2) an additional factor of {\bf M} $\cdot$ {\bf k$_f$} due to the polarization dependence of the RMXS cross section in vertical scattering geometry\cite{Hill_AC_1996}.  The resulting theoretical intensity ratio, I$_{canted}$/I$_{collinear}$ = 0.77, agrees very well with the observed difference in magnetic peak intensities at intermediate temperatures (I$_{FC}$/I$_{ZFC}$ = 0.71 $\pm$ 0.08).

Given this proposed sequence of spin reorientation transitions, it is somewhat surprising that no additional low temperature anomalies are observed in the zfc magnetization data.  This may be explained by the relatively short zfc magnetic correlation lengths determined in Section IIIC.  If the magnetic order in the zfc state is predominantly short-ranged, with many small coexisting magnetic domains, then the resulting magnetization curve could be largely featureless (as observed in Figure 3(a)).  The large enhancement of M$_{ab}$ under fc conditions would then simply be a consequence of the applied field encouraging the formation of larger magnetic domains within the sample, since it will be energetically favorable for the net ferromagnetic moment from the canted spins in each domain to align with the direction of the applied field.

The relative complexity of the fc/zfc magnetic phase diagram of Sr$_3$Ir$_2$O$_7$ is likely a consequence of the interplay between three sets of competing interactions: conventional isotropic Heisenberg superexchange, anisotropic Dzyaloshinskii-Moriya (DM) interactions, and pseudodipolar (PD) interactions associated with the j$_{eff}$ = 1/2 state.  While the detailed properties of these interactions can be quite complex, in general the isotropic superexchange interactions will favor the formation of a collinear ordered state, while the anisotropic DM interactions will favor a canted state.  The interlayer PD coupling is believed to favor a collinear state, while the effect of the intralayer PD term depends on the tetragonal distortion of the IrO$_6$ octahedra (favoring canted order for axial compression, and collinear order for axial elongation)\cite{JWKim_arXiv_2012,Private_Communication}.  The balance between these interactions may be altered by the presence of applied field or by simple structural changes as a function of temperature.  In general, Sr$_3$Ir$_2$O$_7$ exhibits a stronger tendency for collinear ordering because the Ir ions in neighboring bilayers are stacked directly on top of each other.  In addition, because the direction of the octahedral rotations alternates between adjacent Ir-O layers, the DM interactions within this compound are inherently frustrated.

X-ray structure refinements performed on our single crystal samples of Sr$_3$Ir$_2$O$_7$ indicate that the magnitude of the IrO$_6$ octahedral rotations changes from $\sim$ 11.75$^{\circ}$ at room temperature, to $\sim$ 12.52$^{\circ}$ at T = 90 K.  This increased octahedral rotation is likely to enhance the magnitude of the DM interactions which favor the canted AF state.  Sr$_3$Ir$_2$O$_7$ has also been found to exhibit anisotropic thermal expansion.  As the system is cooled from room temperature to T = 90 K, structural refinements indicate that the {\it a}-lattice constant shrinks by $\Delta$a/a = 1.6 * 10$^{-3}$, while the {\it c}-lattice constant expands by $\Delta$c/c = 0.8 * 10$^{-3}$.  This result is supported by recent lattice parameter measurements performed using neutron scattering techniques\cite{Dhital_arXiv_2012}.  This anisotropic lattice expansion may also alter the shape of the IrO$_6$ octahedra, with the larger {\it c}/{\it a} ratio leading to a greater elongation along the axial direction.  As the tetragonal distortion is intimately associated with the intralayer PD interactions, this type of structural change would be likely to favor an increased tendency towards collinear magnetic order\cite{JWKim_arXiv_2012}.  

Similar physics has also been observed in the distorted perovskite YVO$_3$\cite{Ren_Nature_1998,Blake_PRB_2002}.  Like Sr$_3$Ir$_2$O$_7$, this compound undergoes a series of magnetization reversals and spin reorientation transitions upon cooling.  One of these magnetization reversals (at T$_s$ $\sim$ 77 K) is driven by a change in orbital ordering which results in a different Jahn-Teller structural distortion that favors a new spin configuration\cite{Blake_PRB_2002}.  However, the other reversal (at T* $\sim$ 95 K) has been attributed to changes in canting angle driven by the competition between conventional Heisenberg superexchange (which favors collinear AF order), DM interactions, and single ion anisotropy (both of which favor canted AF order, although not necessarily with the same ordering wave vector)\cite{Ren_Nature_1998}.  Although the role of orbital ordering is certainly different in these two systems, this situation appears to offer strong parallels to the case of Sr$_3$Ir$_2$O$_7$, where a shifting balance between three sets of competing interactions as a function of temperature also results in an unusual series of spin reorientation transitions.

\section{Conclusions}

In summary, we have carried out a series of detailed resonant magnetic x-ray scattering measurements on single crystal Sr$_3$Ir$_2$O$_7$.  In particular, these measurements have focused on the evolution of the magnetic Bragg peaks as a function of temperature and applied field.  This work is complementary to previous magnetization studies on this compound performed by Cao et al\cite{Cao_PRB_2002}.  Our RMXS results indicate that Sr$_3$Ir$_2$O$_7$ displays interesting field-induced magnetic behavior, particularly when field-cooled in the presence of modest applied fields (0 $\le$ H $\le$ 0.25 T).  We find evidence of magnetic Bragg peaks along the [1/2, 1/2, L] direction in reciprocal space, consistent with the formation of G-type antiferromagnetic order with multiple magnetic domains.  These magnetic Bragg peaks develop at T* $\sim$ 260 K under both zfc and fc conditions.  Interestingly, this transition temperature is significantly below the first magnetic transition (T$_C$ = 285 K) which is identified in bulk characterization measurements\cite{Cao_PRB_2002, Nagai_JPCM_2007, Nagai_JLTP_2003, Boseggia_arXiv_2012}.  

We propose that the major features of the H-T phase diagram for Sr$_3$Ir$_2$O$_7$ can be understood in terms of competition between two magnetic ground states characterized by collinear AF and canted AF spin configurations.  These two states can be distinguished by the magnitude of the antiferromagnetic order parameter, m$_{AF}$, which is larger for the collinear AF state than the canted AF state.  Under zfc conditions, Sr$_3$Ir$_2$O$_7$ appears to favor the collinear AF state from T* down to T$_D$ $\sim$ 50 K.  Below T$_D$, the system undergoes a spin reorientation transition and adopts a canted AF spin configuration.  Below T $\sim$ 30 K, the ordered structure appears to change again, reverting to a collinear AF state which persists down to the lowest measured temperature.  Under fc conditions, it is the canted AF state which is stabilized from T* down to T$_D$.  At T$\sim$ 30 K the fc system appears to undergo a similar spin reorientation transition, and the canted state is replaced by a collinear AF state that persists down to base temperature.  

In addition to influencing the magnetic phase diagram of Sr$_3$Ir$_2$O$_7$, field-cooling in modest applied fields also encourages the formation of large magnetic domains, and leads to a significant enhancement of magnetic correlation lengths within the {\it ab}-plane.  The effect of higher fields was investigated by measuring the field dependence of magnetic Bragg peaks in applied fields of up to H = 4 T along the [1$\bar{1}$0] direction.  Unlike the single-layered iridate Sr$_2$IrO$_4$, which undergoes a spin-flop transition at H$_C$ $\sim$ 0.2 T [Ref. 15], Sr$_3$Ir$_2$O$_7$ does not appear to exhibit any field-induced phase transitions for H $\le$ 4 T.  We hope that these results will help to stimulate future theoretical and experimental work on Sr$_3$Ir$_2$O$_7$, as well as the other members of the Sr$_{n+1}$Ir$_n$O$_{3n+1}$ series.  In particular, we believe there is now significant motivation for detailed magnetic and structural refinements on this compound as a function of temperature and applied field.  It is important to clarify the nature of the magnetically ordered state between T* and T$_C$, as well as to confirm the orientation of the magnetic moments in the canted and collinear AF states.  Future studies will present intriguing opportunities to explore the effects of alternative field orientations (such as [100] and [001]) and the role of perturbations such as applied pressure and doping.   

\begin{acknowledgments}

The authors would like to acknowledge valuable discussions with H. Gretarsson, J.M. Carter, H.Y. Kee, and G.A. Sawatzky.  Additional support for these measurements was provided by S. LaMarra at X21 and D. Gagliano at 6-ID-B.  Work at the University of Toronto was funded by NSERC of Canada, the Banting Postdoctoral Fellowship program, and the Canada Research Chair program.  Work at the University of Kentucky was supported by NSF through grants DMR-0856234 and EPS-0814194.  Use of the National Synchrotron Light Source at Brookhaven National Laboratory is supported by the U.S. Department of Energy, Office of Science, Office of Basic Energy Sciences, under Contract No. DE-AC02-98CH10886.  Use of the Advanced Photon Source at Argonne National Laboratory is supported by the U.S. Department of Energy, Office of Science, Office of Basic Energy Sciences, under Contract No. DE-AC02-06CH11357.

\end{acknowledgments}

\end{document}